\documentclass[aps,prl,preprint]{revtex4}
\usepackage{CJK}
\usepackage{algorithm,algorithmic}
\usepackage[titletoc,title]{appendix}
\usepackage{mdframed}
\usepackage{subfigure}
\usepackage{float}
\usepackage{wrapfig,blindtext}
\usepackage{graphicx} 
\usepackage{epstopdf} 
\usepackage{pslatex} 
\usepackage{amsmath,amssymb}
\usepackage{caption}
\usepackage{amsfonts,amsthm} 
\usepackage[english]{babel} 
\usepackage[dvipsnames]{xcolor}
\usepackage{wasysym}
\usepackage{pdfpages}
\usepackage{float}
\usepackage{comment}
\usepackage{soul}

\usepackage{nomencl}
\usepackage{bm}
\usepackage{threeparttable}
\usepackage{makecell}
\usepackage[makeroom]{cancel}

\usepackage[font=small,labelfont=bf,
   justification=justified,format=plain]{caption}

\usepackage{comment}

\usepackage{nomencl}
  \makenomenclature

\begin{document}
\begin{CJK*}{GB}{}

\title{Perspectives on Machine Learning-augmented Reynolds-averaged and Large Eddy Simulation Models of Turbulence}

\author{Karthik Duraisamy}

\address{Department of Aerospace Engineering, University of Michigan, Ann Arbor, Michigan 48109, USA}

\begin{abstract}
	
This work presents a review and perspectives on recent developments in the use of machine learning (ML) to augment Reynolds-averaged Navier--Stokes (RANS) and Large Eddy Simulation (LES) models of turbulent flows.  Different approaches of applying supervised learning to represent unclosed terms, model discrepancies and sub-filter scales are discussed in the context of RANS and LES modeling.  Particular emphasis is placed on the impact of the training procedure on the consistency of ML augmentations with the underlying physical model.  Techniques to promote model-consistent training, and to avoid the requirement of full fields of direct numerical simulation data are detailed.  This is followed by a discussion of physics-informed and  mathematical considerations on the choice of the feature space, and imposition of constraints on the ML model.  With a view towards developing generalizable ML-augmented RANS and LES models, outstanding challenges are discussed, and perspectives are provided. While the promise of ML-augmented turbulence modeling is clear, and successes have been demonstrated in isolated scenarios, a general consensus of this paper is that truly generalizable models require model-consistent training with  careful characterization of underlying  assumptions and imposition of physically and mathematically informed priors and constraints to account for the inevitable shortage of data relevant to predictions of interest. Thus, machine learning should be viewed as one tool in the turbulence modeler's toolkit. This modeling endeavor requires multi-disciplinary advances, and thus the target audience for this paper is the fluid mechanics community, as well as the computational science and machine learning communities. 

\end{abstract}

\maketitle
\end{CJK*}
\section{Introduction}
 Over the past decade, with the availability of rich datasets from direct numerical simulations (DNS) and experiments, there has been a rapid growth in the use of machine learning methods in fluid mechanics~\cite{brunton2020machine}. The emergence of data science as a discipline in its own right, and broad accessibility of essential machine learning tools has fostered these advances. While rapid progress has been made in {\em data-driven} methods in areas such as flow decomposition~\cite{taira2020modal}, governing equation discovery~\cite{rudy2019data} (for simple systems), and reinforcement learning~\cite{verma2018efficient} for flow control, predictive modeling of turbulent flow has followed a {\em data-enabled} paradigm. In other words, existing modeling knowledge is leveraged, and data is used to {\em improve } existing model structures. This is mainly because of the requirement of simulation models to be applicable across different flow regimes, geometric configurations, and the need to incorporate boundary conditions, all of which necessitate a coarse-grained partial differential equation setting. This paper is thus specifically focused on recent developments in the use of machine learning to {\em augment} Reynolds-averaged Navier--Stokes (RANS) and Large Eddy Simulation (LES) models. While it is well-argued that data has always played an important role in the development of models of turbulent flows, the past decade has witnessed attempts towards a more comprehensive and formal use of data in model development~\cite{duraisamy2019turbulence}. 
 
  The earliest application of formal machine learning in fluid mechanics is hard to ascertain. Teo et al.'s~\cite{teo1991neural} (1991)  application to particle tracking and Milano et al.'s~\cite{milano2002neural} (2002) application to near-wall reconstruction  appear to be among the first explorations in experimental and computational fluid dynamics, respectively. In the context of predictive turbulence modeling, the use of ML has its origins in the uptick in research activities in uncertainty quantification in physical modeling 15 years ago. A comprehensive review of uncertainty quantification (UQ) in RANS  modeling can be found in Ref.~\cite{xiao2019quantification}. A notable UQ work from this era, and one that is relevant to the present context is that of Oliver \& Moser~\citep{oliver2011bayesian} (2011), who introduced a discrepancy to the Reynolds stress predicted by a baseline RANS model, in the form of a spatially-dependent Gaussian random field. Dow \& Wang~\cite{dow2011quantification} (2011) applied a similar approach, augmenting the eddy-viscosity  rather than the Reynolds stress. In contrast to the representation of Oliver \& Moser, however, Dow \& Wang assimilated data from multiple geometries rather than from a single geometry. Though these papers do not  involve ML, they  are particularly relevant to the emergence of ML applications, because they address model inadequacies in a non-parametric fashion, in contrast to earlier work which was focused on parameter calibration. 
  
  The first wave of ML applications in RANS modeling can be tracked to Tracey et al.~\cite{tracey2013application} (2013). In this work, supervised learning was performed in an a priori setting on DNS data, with the goal of reconstructing discrepancies in the Reynolds stress anisotropy tensor. The key advance from the UQ work was the transformation of model discrepancies from the spatial domain $x$ to a feature space $\eta(x)$, consisting of  quantities such as mean velocity gradients. 
  Vollant et al.~\cite{vollant2014optimal} (2014) applied similar feature space learning in an optimal estimation setting for subgrid scale modeling for LES.  Further advances in enforcing consistency between the data and the model~\cite{duraisamy2015new,parish2016paradigm}, embedded invariance~\cite{ling2016machine}, symbolic regression~\cite{weatheritt2016novel}, etc. characterize the first phase of research. In other early work in the field, Ma et al.\cite{ma2015using,ma2016using} (2016) used neural networks to model the inter-phase mass and momentum fluxes in multiphase flow simulations.  Since 2016, this field has exploded in activity as a large number of researchers have pursued ML augmentation and applications have expanded to LES, multiphase flows, and combustion modeling.

  Section~\ref{sec:problem} establishes the problem statement of coarse-graining and closure modeling, and   is presented in a form that is also accessible to the computational science community. Section~\ref{sec:representation} presents the ways in which model discrepancies have been represented via supervised learning algorithms. This is the area that has seen the highest level of activity in the literature. Sec.~\ref{sec:consistency} studies the process by which the models are trained. Particularly, the concept and importance of achieving consistency between the learning and prediction environments is emphasized.  Section~\ref{sec:features} details general principles relevant to the selection of an appropriate feature space.  Section~\ref{sec:constraints} discusses ways in which physics-informed constraints can be placed on the ML model and ML-augmented models. Section~\ref{sec:perspectives} presents additional challenges and perspectives towards the goal of developing reliable and robust  ML-augmented models for turbulence. 

\section{Problem statement}
\label{sec:problem}
Represent the highest fidelity simulation (e.g. Navier--Stokes equations, as solved by a direct numerical simulation) in the form
\begin{equation}
 \mathcal{R} (q) = 0,
\label{eq:truth}
\end{equation}
where $\mathcal{R}$ is a PDE operator, and $q$ are the state variables (pressure, velocity, energy, etc.). In  practical problems - such in an airplane wing or a gas turbine combustor - the level of spatial and temporal resolution required is so high that direct simulations will be unaffordable for decades to come. Thus, reduced-fidelity models  are sought.  In reduced-fidelity modeling using RANS \& LES, the state variables are decomposed into coarse-grained variables $\tilde{q}$ and unresolved variables $\hat{q} = q-\tilde{q}$. For example, in LES, the coarse-graining operation may represent implicit or explicit low-pass filtering, and in RANS, it may represent ensemble averaging.   Applying such a coarse-graining operation  to  $\mathcal{R}(q)$ yields
\begin{equation}
 \widetilde{\mathcal{R}(q)} = 0 \rightarrow  \mathcal{R}(\tilde{q})+\mathcal{N}({F}(q)) = 0.
\label{eq:cg}
\end{equation}
It is notable that Eq.~\ref{eq:cg} is {\em not} an approximation. While the form of the operator $\mathcal{N}$ is known, and the quantity $F$ is well-defined, the latter is {\em not closed} in the coarse-grained variables $\tilde{q}$.  
For instance, in incompressible single phase flow, the Navier--Stokes equations are given by
$$
    \frac{\partial u}{\partial t} + u \cdot \nabla u = -\frac{1}{\rho}\nabla p + \nu \nabla^2 u,
$$
where the uknowns are $q \equiv \{u, p\}$ (the velocity and pressure), and $\nu, \rho$ represents the kinematic viscosity and density. When ensemble averaging $\tilde{\cdot}$ is introduced and applied  to the above equation, we get
$$
\frac{\partial \tilde{u}}{\partial t} + \tilde{u} \cdot \nabla \tilde{u} = -\frac{1}{\rho}\nabla \tilde{p} + \nu \nabla^2 {\tilde u} - \nabla \cdot (\widetilde{\hat{u} \otimes \hat{u}}).
$$
These are the RANS equations and thus, $\mathcal{N}() \equiv -\nabla \cdot ()$ and ${F}(q) \equiv \widetilde{\hat{q} \otimes \hat{q}}$. In this work, the quantity $F$ is used to generically represent unclosed terms such as the Reynolds stresses,  subgrid scale stresses, scalar fluxes, turbulence-chemistry interactions, etc.


The goal of turbulence modeling is to overcome the closure problem in Eq.~\ref{eq:cg} by defining a RANS or LES model in the form
\begin{equation}
 \mathcal{R}(\tilde{q}_m)+\mathcal{N}(F_m(\tilde{q}_m,\tilde{s}_m)) = 0.
\label{eq:model}
\end{equation}
The main objective in turbulence modeling is to construct an closure approximation $F_m \approx F$  in terms of the modeled coarse-grained variables $\tilde{q}_m$  and some secondary variables $\tilde{s}_m$, which may themselves involve additional transport equations $\mathcal{G}_m(\tilde{s}_m,\tilde{q}_m) =0.$ For example, $\tilde{q}_m$ can represent the modeled ensemble averaged velocity and $\tilde{s}_m$ can represent the modeled dissipation rate of turbulent kinetic energy as shown in Table ~\ref{tableclosure}.

The construction of $F_m$ and $\mathcal{G}_m$ is a highly intricate process, evolving over several decades through a combination of  physical insight, mathematics and empiricism~\cite{durbin2017some}. While data has always been an enabler of closure modeling, the ways in which data has been used thus far has perhaps not been comprehensive. Partly as a consequence of this, many closure modeling approaches have saturated in accuracy ~\cite{spalart2015philosophies,duraisamy2017status} over the past decade (or more). 

In this review, we specifically discuss the use of ML in augmenting coarse-grained RANS and LES models, and  will use $\tilde{q}, \tilde{s}$ to denote {\em coarse-grained} quantities derived from a DNS or experiment, and $\tilde{q}_m,\tilde{s}_m$ will be modeled representations of the same quantities. $Y(\tilde{q}),Y(\tilde{q}_m)$ will be used to represent observables of the resolved field (e.g. skin-friction coefficient on a wall). The machine learning model will be generically represented by $\delta_m(\tilde{\eta}_m; w)$, where $\tilde{\eta}_m$ are features (inputs) and $w$ are the parameters of the learning representation (e.g. weights and biases of a neural network). For illustrative purposes, examples of the above quantities are shown in Table~\ref{tableclosure}. The machine learning representation $\delta_m$ can represent the entire closure term (for instance, Beck et al.~\cite{beck2019deep} develop an ML model for the subgrid scale stresses in LES), or part of a closure term (for instance, Schmelzer et al.~\cite{schmelzer2020discovery}) extract an ML model as a correction term to an eddy viscosity model of the reynolds stress tensor in RANS).

\begin{table}[!ht]
	\begin{center}
		\begin{threeparttable}
			\caption {Examples of closure terms and related quantities}\label{tableclosure}
			\begin{tabular}{|c|c|c|c|c|c|}
				\hline
		Closure Term & $F$ & $\tilde{q}_m$ & $\tilde{s}_m$ & $\tilde{\eta}_m$  \\ \hline \hline
	 Reynolds stress (RANS) & $\tau = \widetilde{(\hat{u} \otimes \hat{u})}$ &$\tilde{u}_m$ & $k_m,\epsilon_m$ & $\tilde{S}_m\frac{k_m}{\epsilon_m}$ \\		\hline
		Subgrid scalar flux (LES)	&   $f=\widetilde{u T} - \tilde{u}\widetilde{T}$ &$\tilde{u}_m, \tilde{T}_m$ & $ - $ & $\Delta^2 |\tilde{S}_m|  \nabla \widetilde{T}$ \\		
				\hline
			\end{tabular}
		 \begin{tablenotes}
			\item[] $u_m,T_m$ : Velocity, Temperature; $k_m,\epsilon_m$ : Turbulent kinetic energy \& dissipation rate; $S_m$ : Strain-rate; $\Delta$ : Filter size.
		\end{tablenotes}
		\end{threeparttable}
	\end{center}  
\end{table}

\section{Representation}
\label{sec:representation}
Much of the research in ML-augmented turbulence modeling has gone towards different ways of representing model inadequacy and embedding them appropriately in RANS and LES models. It is not the intent of this section to cover all of the work in this area, but rather an attempt is made at presenting a prototypical set of representations.  This section is divided into two parts: The first discusses techniques to model the impact of the unresolved physics on the coarse-grained variables $F_m(\tilde{q}_m,\tilde{s}_m)$, and the second part discusses techniques to model the unresolved quantities $\hat{q}_m$ directly.


\subsection{Closure term \& model inadequacy representations}

Tracey et al.~\cite{tracey2013application} represened the Reynolds stress tensor as   $$\tau_m = 2 k_m \left(\frac{1}{3}I + \tilde{V}_m (\tilde{\Lambda}_m + \delta_m(\tilde{\eta}_m; w) \tilde{V}_m^T) \right),$$  where $\tilde{k}_m$ is the modeled turbulent kinetic energy,  
$\tilde{V}_m,\tilde{\Lambda}_m$ are matrices that contain the eigenvectors and eigenvalues of the modeled Reynolds stress anisotropy, respectively.  $\delta_m(\tilde{\eta}_m; w)$ is a Kernel regression-based ML model that models the error in anisotropy via supervised learning on a DNS dataset. The features $\tilde{\eta}_m$ include the eigenvalues of the anisotropy tensor, the ratio of the production-to-dissipation rate (of turbulent kinetic energy) and a marker function that masks regions of thin shear layers from augmentation. 
 In a series of papers~\cite{wang2017physics-informed,wu2018physics}, Xiao and co-workers expanded the description above to include a more comprehensive perturbation of not just the eigenvalues, but also the eigenvectors and the turbulent kinetic energy, and considered a broad range of features. Duraisamy et al.~\cite{duraisamy2014transition,duraisamy2015new,singh2017machine-learning-augmented} applied similar feature-based augmentations to transport equations, rather than directly to the Reynolds stress.

Ling and Templeton~\cite{ling2016reynolds} proposed a neural network architecture to learn  the coefficients of a tensor basis expansion for the Reynolds stresses in the form 
 \begin{equation}
\label{eq:astm}
\begin{aligned}
\tau_m =  2 k_m   \left(\frac{1}{3}I + \left[\sum_{n=1}^{10}\delta_m^{(n)}(\tilde{\eta}_m;w){T}^{(n)}(\tilde{S}_m,\tilde{\Omega}_m) \right] \right),
\end{aligned}
\end{equation}
where ${T}^{(n)}(\tilde{S}_m,\tilde{\Omega}_m)$ are the tensorial basis~\cite{pope2000turbulent} (with respect to the strain rate and vorticity tensors) and $\delta_m^{(n)}(\tilde{\eta}_m;w)$ are the coefficients which are represented using neural networks. It is notable that the features $\tilde{\eta}_m$ are taken to be five invariants based on $\tilde{S}_m$ and $\tilde{\Omega}_m$, and are objective by definition.  This so-called tensor basis neural network has since been extended in application to model turbulent heat fluxes~\cite{milani2020generalization} and scalar fluxes~\cite{milani2020turbulent}. Kaandorp and Dwight~\cite{kdp} leverage the above integrity basis, but use random forest regression instead of neural networks.

Approaches that are similar in spirit - from the viewpoint of constructing models for the Reynolds stress anisotropy in terms of invariants of the velocity gradient tensor - are pursued in \cite{weatheritt2016novel,weatheritt2017development} using symbolic regression  with genetic programming. Symbolic regression has the appeal of explicit and interpretable model forms that are more amenable for analysis. Ref.~\cite{haghiri2020large} extends the above invariant-based gene-expression approach to represent turbulent diffusivity.

Schmelzer et al.~\cite{schmelzer2020discovery} employ sparse linear regression (rather than neural networks or symbolic regression)  on a library of candidate functions which are written as tensor polynomials of the aforementioned invariants.  Beetham et al.~\cite{beetham2020formulating} also follow a similar approach and further  extend sparse linear regression to model drag production, drag exchange, pressure strain, and viscous dissipation in RANS of multiphase flows~\cite{beetham2020sparse}. Sparsity-enforcing regularizers in these approaches lead to the  elimination of some model terms, thus simplifying the model form.
  
Sarghini et al.~\cite{sarghini2003neural} introduced the idea of using neural networks to aid  subgrid scale modeling in LES. The feature space included the resolved velocity gradients and stresses, and the output was a Smagorinsky-stype viscosity coefficient. More recently, several researchers (e.g. ~\cite{gamahara2017searching,zhou2019subgrid,pawar2020priori}) have used similar approaches to directly relate the subgrid scale stress tensor to a feature space which includes the resolved velocity gradient tensor and mesh resolution. Notable work was performed by Vollant et al.~\cite{vollant2014optimal, vollant2017subgrid-scale}, who used optimal estimation theory to separate subgrid scale modeling error into parametric and model form components and  neural networks to address each of the components in isolation. 

 Beck et al.~\cite{beck2019deep} (using convolutional neural networks on the resolved field), and Xie.  et al.~\cite{xie2019modeling,xie2020modeling} (using the resolved local flow gradients ) directly model the subgrid scale forcing $f \approx \nabla \cdot \tau_m$, since small errors in the subgrid scale {\em stresses} can otherwise generate large errors in the momentum equation. 


  Over the past year, similar approaches to subgrid scale modeling have also been applied to reacting flows and multiphase flows. As examples, Ref.~\cite{lapeyre2019training} uses convolutional neural networks on the full field of the resolved progress variable to construct subgrid flame density function estimates; and Ref.~\cite{yellapantula2020machine} uses  neural networks to learn the filtered progress variable source term.





\subsection{Subgrid scale representations}
The idea of approximate deconvolution as a path towards closure modeling was introduced by Stolz \& Adams~\cite{stolz1999approximate} . The basic idea revolves around reconstructing sub-filter contributions $\hat{q}_m$ from filtered quantities $\tilde{q}_m$. Once such an approximation is constructed, the sub-grid scale stresses can be approximated as $\tau_m = \tilde{q}_m\otimes\tilde{q}_m - \widetilde{{q}_m \otimes {q}_m}$. While the original idea of approximate deconvolution was based on analytically inverting a known filtering kernel, \cite{maulik2017neural,yuan2020modeling} employ a neural network to represent the deconvolution $\hat{q}_m = \delta_m(\tilde{q}_m; w)$, based on a localized stencil of inputs. Note that unlike in the classical approximate deconvolution strategy, the filtering kernel is not assumed a priori.

A related idea to deconvolution is super-resolution. While deconvolution is aimed at modeling sub-filter quantities on the same discretization as the resolved quantities, super-resolution targets the extraction of a finer resolution field - for instance, extracting finer mesh quantities or a higher-order accurate solution.  Inspired by advances in imaging, \cite{xie2018tempogan,fukami2019super} introduced the idea of super-resolution to fluid mechanics by leveraging neural networks to represent the mapping from the coarse field to the fine field.
\cite{fukami2020machine} extended applications to the spatio-temporal setting, where  given a coarse field at two time instances $t$ and $t+\Delta t$,  super-resolved spatio-temporal fields are generated at  many smaller time instances in the interval $[t..t+\Delta t]$. Refs.~\cite{subramaniam2020turbulence} and ~\cite{kim2020unsupervised} use variants of Generative Adversarial Networks (GAN) for super-resolution. Though these models have mostly been evaluated in an a priori sense, these approaches are highly expressive, and present promise for online modeling.

In Ref.~\cite{bode2019using}, a GAN is trained to super-resolve a coarse solution (filtered DNS) to a fine solution (DNS). The resolution (image pixels) of input and output are kept fixed. The model is used to obtain fine-scales from the coarse LES solution, which is  used to compute the subgrid terms (in the momentum and scalar equations). 

DNS of dispersed multiphase flow is expensive since the field around each particle needs to be resolved leading to a large number of variables. Ref.~\cite{siddani2020machine} uses a GAN to generate the velocity field around a group of particles in a small section of the domain locally, which is then applied patchwise over the entire domain. 

\section{Training \& Consistency}
\label{sec:consistency}
The previous section was focused on the {\em representation} of the unclosed terms, i.e. on how to specify the model form of the closure. In this section, we study the process by which these models are extracted from data. The concept of consistency between the learning and prediction environments is explored. To simplify notations, we will use $\delta_m(\tilde{\eta}_m; w)$ to represent the ML models, and   $\mathcal{R}_a(\tilde{q}_m,\tilde{s}_m,\delta_m(\tilde{\eta}_m;w)) = 0$ to represent all of the transport equations (e.g. Reynolds-averaged mass, momentum, energy equations and transport equations for scale-providing variables) of the machine learning augmented model. Such a definition is used to exemplify the representations in Section~\ref{sec:representation} in a compact manner.

\subsection{A priori training}
In a majority of the references above (e.g. ~\cite{tracey2013application,wu2018physics,ling2016machine,weatheritt2016novel,beetham2020sparse}), training is performed by directly extracting $\delta$ (the target of the learning model) and $\tilde{\eta}$ (the input to the learning model) from the DNS. Following this, one posits a ML model $\delta_m(\cdot; w)$, and the  following supervised learning problem is posed:

\begin{equation}
\min_w \mathcal{L}[\delta, \delta_m(\tilde{\eta}; w)],
\label{eq:loss_apriori}
\end{equation}
where $\mathcal{L}$ is a generic loss function that is a proxy for frequentist or Bayesian inference. 
For instance, in sparse regression,  $$\mathcal{L}[\delta, \delta_m(\tilde{\eta}; w)] \equiv  || \delta - \delta_m(\tilde{\eta}; w)||_2^2 + \lambda ||w||_1.$$

After the training (and cross-validation) process is complete, the trained model $\delta_m(\cdot;w)$ is embedded in a baseline model in a predictive setting:
\begin{equation}
 \mathcal{R}_a(\tilde{q}_m,\tilde{s}_m,\delta_m(\tilde{\eta}_m;w)) = 0.
\label{eq:pred}
\end{equation}

This training approach is natural, non-intrusive (i.e. the solver is not involved), and provides opportunities to directly impose physics-based constraints. However, consistency with the model can become a critical issue as has been pointed out in the context of RANS~\cite{singh2017augmentation,duraisamy2019turbulence,taghizadeh2020turbulence} and LES~\cite{freund2019dpm}. In short,  a priori training establishes the consistency of the ML model with the DNS field, but does not guarantee consistency with the RANS or LES environment. The following are some of the main reasons for the loss of consistency:

$\bullet$ {\em Feature mismatch between training and prediction:}  During the training process (Eq.~\ref{eq:loss_apriori}), the coarse-grained features $\tilde{\eta}$ from the DNS are used as inputs to the learning model, whereas in the prediction process, coarse-grained model features $\tilde{\eta}_m$ are used. Thus, for this approach to work well, the model has to predict the features $\tilde{\eta}_m$ very accurately (i.e. $\tilde{\eta}_m = \tilde{\eta}$). This is typically difficult to achieve because the features may depend on secondary variables which may not be predicted well by the models. For instance, a feature $\tilde{S}\frac{k}{\epsilon}$ may not equal $\tilde{S}_m\frac{k_m}{\epsilon_m}$ because $k_m$,  $\epsilon_m$ can be very different from $k,\epsilon$ even if $\tilde{S}_m$ is close to $\tilde{S}$. This is because models are typically constructed to predict certain quantities (e.g. mean flow, Reynolds shear stress, etc.) well, but not all quantities. This is especially true in RANS where several secondary quantities are used to provide length and time-scales which  should not be interpreted to have a one-on-one correspondence with the DNS quantities.  

\begin{figure}
	\centering
	\subfigure[\ \ Offline performance (a priori training)]{\includegraphics[height=0.45\textheight]{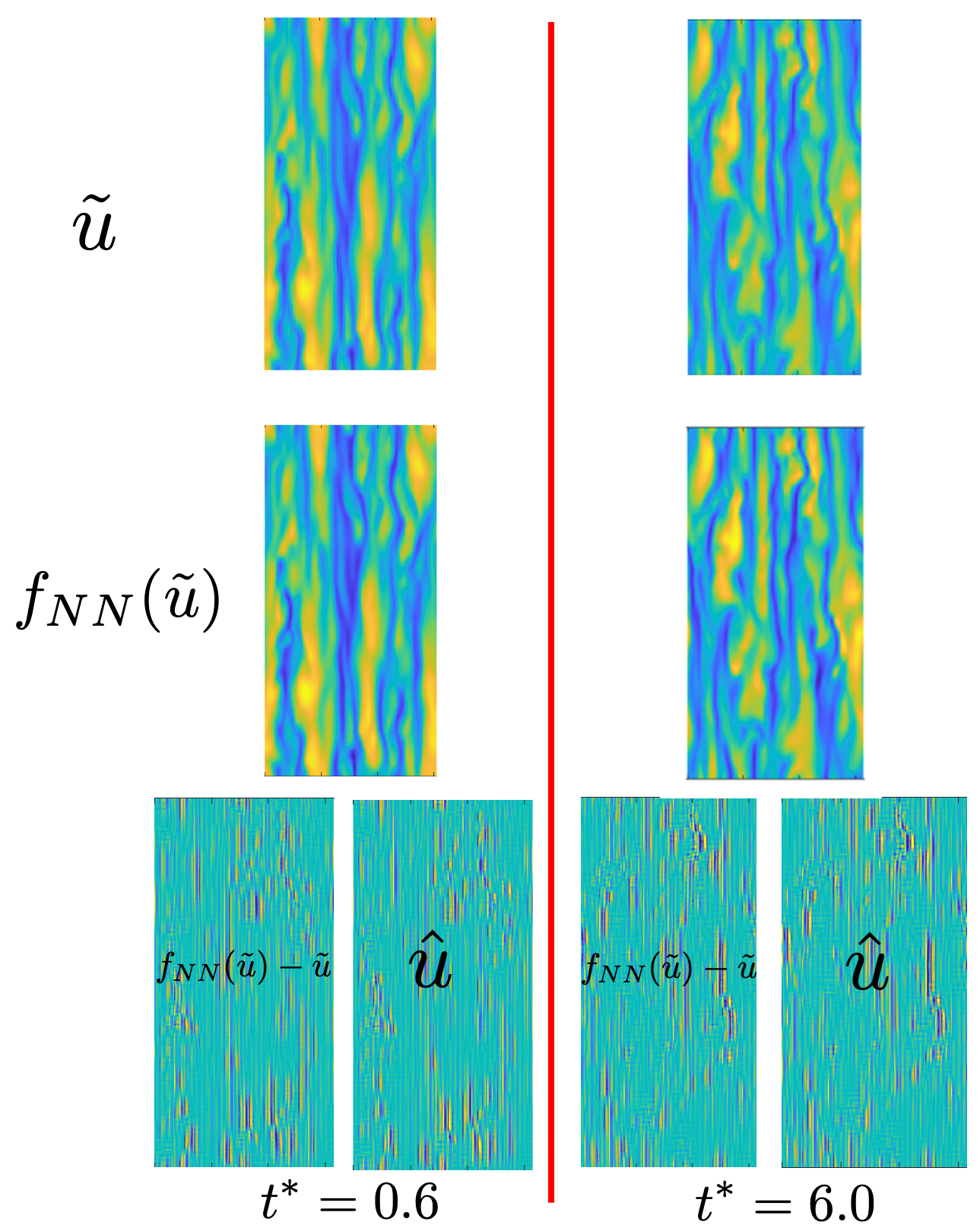}}
	\subfigure[\ \ On-line performance (prediction)]{\includegraphics[height=0.45\textheight]{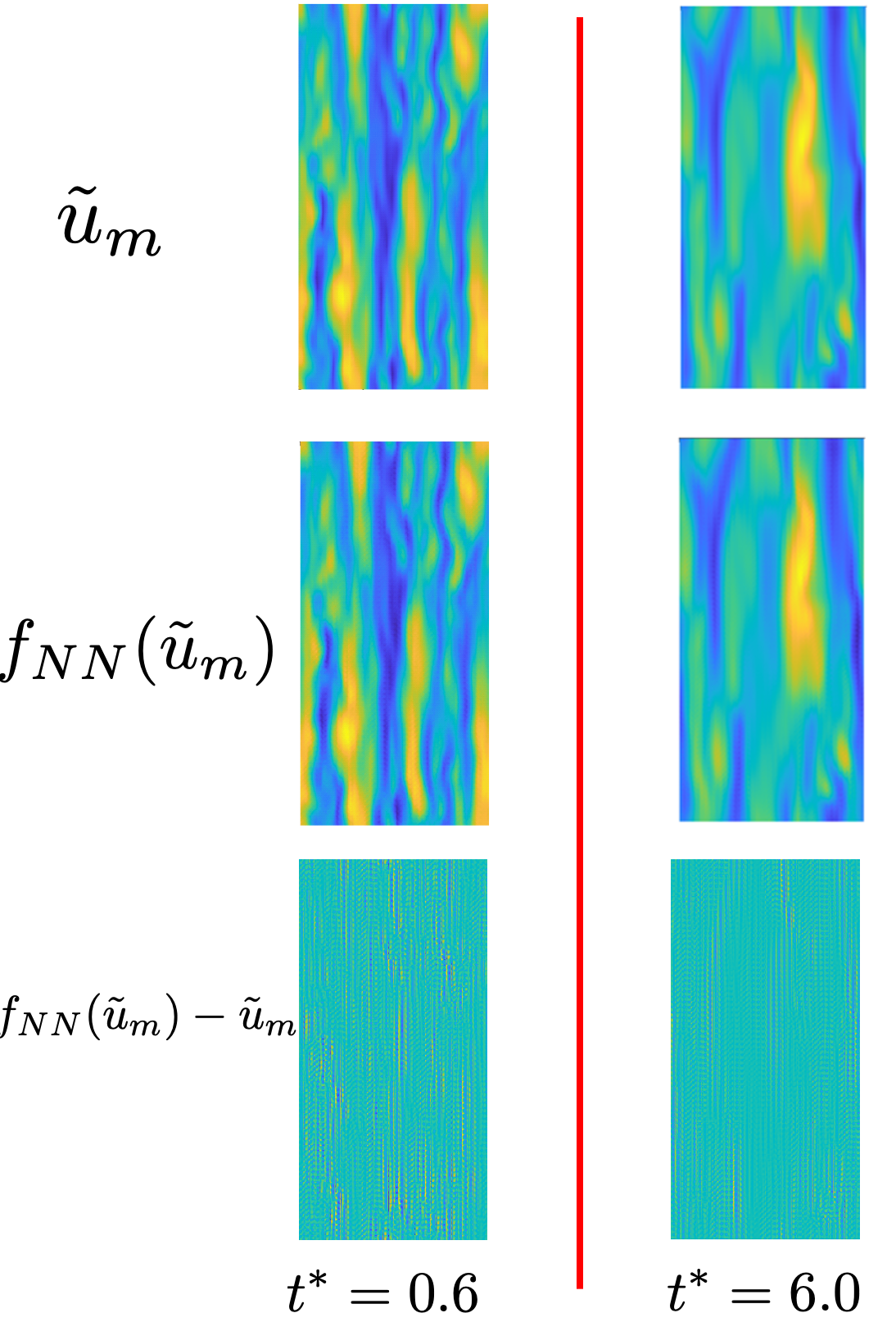}}
	\caption{Impact of error accumulation in online simulations. Top row : coarse-grained field; Middle row : fine field; Bottom row : fine scales.}
	\label{fig:sr}
\end{figure}

$\bullet$ {\em Accumulation of errors in prediction:} Consider a super-resolution operation of going to a fine field from a coarse field via a neural network. The results of such a process is shown in Figure~\ref{fig:sr}.  The neural network is trained (offline) to represent $u(t) = f_{NN}(\tilde{u}(t); w)$, whereas when it is embedded in the solver, it predicts $u_m(t) = f_{NN}(\tilde{u}_m(t); w) = f_{NN}(\tilde{u}(t)+e(t); w)$. As the error $e(t)$ accumulates over time, the neural network is required to make predictions based on a field that is corrupted by error, which becomes futile as seen by the lack of fine-scale structure at $t^\ast=6.0$ in the on-line prediction. Such errors are typical of most practical LES computations because even though one attempts to model only the unresolved scales, the scales that are barely larger than the filter size are often very poorly resolved. Thus, the learning model has to consider the corruption of the resolved scales.

$\bullet$ {\em Balance between model terms:} It is well-recognized in the turbulence modeling community that successful a priori evaluation is neither a necessary (e.g. Smagorinsky model) nor a sufficient condition for successful predictive models. Even second moment closure-based RANS models are formulated  such that the balance between model terms is managed in a manner that ensures a degree of success in predictive outcomes.  \cite{raiesi2011evaluation} calculated the working  variables in  turbulence models using DNS and LES datasets for one and two-equation models and concluded that the use of exact values of the turbulent kinetic energy and dissipation rate in the modeled eddy-viscosity did not improve its performance. Ref.~\cite{thompson2016methodology} showed that even substituting Reynolds stress fields from reputable DNS databases may not lead to satisfactory velocity fields. 
Further, ~\cite{wu2018on} investigated potential conditioning problems that arise when explicitly trained ML models are injected into existing turbulence models. 

Another major impediment to the use of a priori techniques is that a {\em full field of DNS data is required} to train the model. Since DNS data will not be available in practical regimes, this is a major limitation.

In spite of the above challenges, good results have been reported in the literature. For the purposes of generalization, it will be a good practice to ascertain the correlations between $\tilde{\eta}$ \& $\tilde{\eta}_m$ (beyond $\delta$ \& $\delta_m$) such that the degree of loss in consistency can be monitored.

\subsection{Model-consistent training}

The previous discussion highlights the importance of learning model augmentations in an (imperfect) modeled environment rather than the (perfect) environment of the coarse-grained DNS field. Enforcing consistency, however, requires the solution of inverse problems that minimize a discrepancy between the model output and the data. Thus, the model $R_a(\cdot)$ is involved in the training process in contrast to a a priori training.

Given a sparse dataset $\tilde{Y}^i$ from experiments or DNS,  the so-called field inversion approaches~\cite{dow2011quantification,duraisamy2015new,singh2016using}  seek the spatio-temporal model augmentation field $\delta_m^i(x,t)$ via the following PDE-constrained inverse problem:
\begin{align}
\min_{\delta^i_m}  & \mathcal{L}[{Y}^i,{Y}^i_m(\delta_m^i)], \  \ \textrm{s.t.}   \  \ \mathcal{R}_a(\tilde{q}_m,\tilde{s}_m,\delta_m(\tilde{\eta}_m;w)) = 0,
\label{eq:fi}
 \end{align}
 
 Such an augmentation is consistent with the underlying model, and unlike the a priori approach which requires $\delta^i$ (rather than $Y^i$) from a detailed DNS field, this approach - in principle -  can assimilate information from available and potentially sparse data.  In practice, one has to ensure that the data $Y^i$ is informative of the underlying model discrepancy.
 
 The field inversion (FI) approach has been pursued  in the context of augmentation of eddy viscosity~\cite{dow2011quantification}, transport equation terms~\cite{singh2016using,kohler2020data,yang2020improving},  Reynolds stresses~\cite{xiao2016quantifying,duraisamy2016informing}, and in the mean flow momentum equation~\cite{franceschini2020mean}. Given a dataset $Y^i$, the FI approach yields $\delta^i_m$, which is a spatio-temporal field, and is thus problem-specific. To convert this into a generalizable augmentation, FI is performed on $k$ datasets that are presumably informative of the model discrepancy, following which a set of targets $\delta_m^\ast = \{\delta_m^1,...,\delta_m^k\}$ and features $\tilde{\eta}_m^\ast = \{\tilde{\eta}_m^1,...,\tilde{\eta}_m^k\}$ are collected. Supervised learning can then be performed:
 
\begin{equation}
\min_w   \mathcal{L}[\delta^\ast_m, \delta_m(\tilde{\eta}_m^\ast; w)].
\label{eq:loss_fiml}
\end{equation}
In contrast to Eq.~\ref{eq:loss_apriori} in which coarse-grained DNS data is used to train the model, Eq.~\ref{eq:loss_fiml} creates data from the model outputs to train a generalized augmentation. This combination of FI and ML as given in equations~\ref{eq:fi} and ~\ref{eq:loss_fiml} has been pursued by several researchers ~\cite{parish2016paradigm,singh2017machine-learning-augmented,singh2017augmentation,zhu2019machine,kohler2020data,yang2020improving} in various applications in turbulence and transition modeling. 

While the aforementioned FIML approach promotes consistency between the model outputs and the learning environment, imperfections in the learning process (Eq.~\ref{eq:loss_fiml}) can introduce a degree of inconsistency between the learning and prediction environments. This remaining inconsistency can be eliminated by integrating the inference step with the learning step in the form:  

\begin{align}
\min_w   \mathcal{L}[{Y}, {Y}_m(\delta(\tilde{\eta}_m;w))], \ \ 
\  \ \textrm{s.t.}   \  \  \mathcal{R}_a(\tilde{q}_m,\tilde{s}_m,\delta_m(\tilde{\eta}_m;w)) = 0.
\label{eq:IFIML}
\end{align}

\begin{figure}
	\centering
	\includegraphics[width=0.6\textwidth]{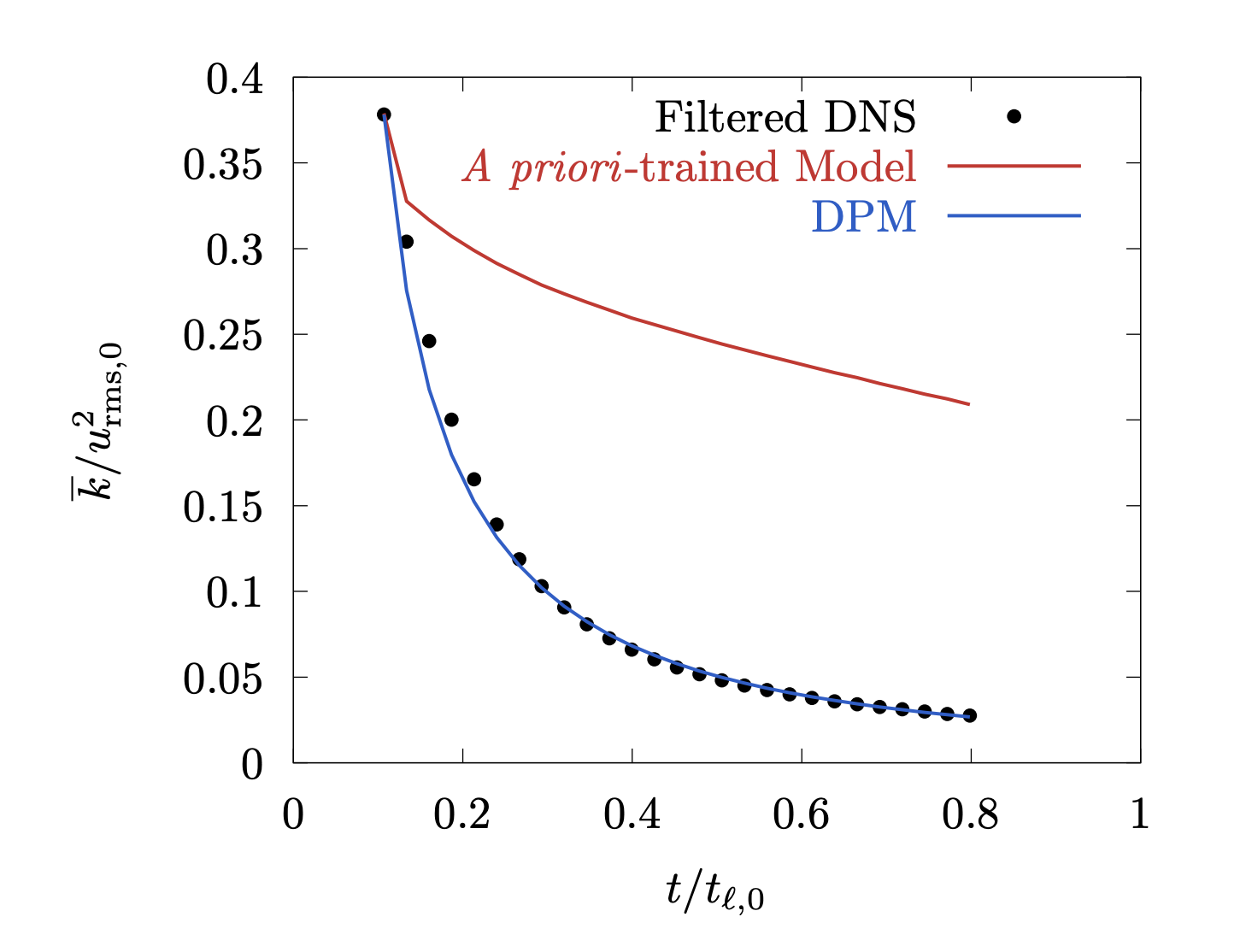}
	\caption{Comparison of apriori-trained subgrid scale closure and model consistent learning/inference (DPM) applied to LES of homogeneous isotropic turbulence (Ref.~\cite{freund2019dpm}) .}
	\label{fig:freund}
\end{figure}

In this approach, the ML output is first fed into the model, and the model output is used in the loss function. Notably, both of these steps are integrated within a single optimization step that updates the weights of the ML model. This tightly coupled approach  has recently been employed in the context of RANS~\cite{holland2019towards,holland2019field} and LES~\cite{freund2019dpm}, and ensures full consistency between the learning and prediction environments. Figure~\ref{fig:freund} exemplifies the importance of consistent training.

It has to be mentioned that one minor drawback of the integrated approach (Eq.~\ref{eq:IFIML}) is that features have to be selected before the inference, whereas separation of the FI (Eq.~\ref{eq:fi}) and ML (Eq.~\ref{eq:loss_fiml}) stages yields more flexibility. 

It is notable that model-consistent training  involves large scale inverse problems, typically necessitating the use of  adjoint-driven~\cite{giles2003algorithm} optimization techniques~\cite{parish2016paradigm,holland2019field,freund2019dpm}. Given the intrusive nature of adjoints, this presents major challenges to development, and has proven to be a barrier  for researchers to develop model-consistent ML augmentations. This is especially a challenge in tightly coupled approaches, as the optimization considers the scale and non-linearities of the PDE and highly non-convex loss function landscapes of neural networks. An additional challenge in LES  is that chaoticity can lead to unstable adjoint solutions that require special treatment~\cite{blonigan2018multiple,ni2019sensitivity}.  

To circumvent the complexity of the afore-mentioned tightly coupled optimization problems, weakly coupled techniques such as `embedded learning'~\cite{holland2019towards}, `iterative machine learning' ~\cite{liu2019iterative} , 'CFD-driven machine learning'~\cite{zhao2020rans}, and `closed loop training'~\cite{taghizadeh2020turbulence} have been proposed. While some of these approaches still require a full field of DNS data as in a priori training, these techniques represent a movement towards establishing model consistency without the need for complex adjoint-driven machinery, thus reducing barriers to development of ML-based turbulence models. 

Convergence characteristics of all the methods mentioned in this section have to be studied in a mathematically rigorous fashion, and will be fruitful research direction. The choice of the loss function is also a topic that has not been studied in earnest.



\section{Feature selection}
\label{sec:features}

 The ML models discussed above are based on regression and are meant to be interpolative in feature space $\eta_m$. 
 However, if features are properly selected, and the feature space is adequately populated, embedding the output of these regression  models within a suitable physics-based model $\mathcal{R}_a(\tilde{q}_m,\tilde{s}_m,\delta_m(\tilde{\eta}_m;w)) = 0$ can yield better predictive properties on unseen geometries and flow configurations. 
 
 In the limit of the availability of an infinite amount of data, feature selection can be posed as an unsupervised learning problem. In a practical turbulence modeling scenario, however,  selection of features should be guided by turbulence modeling principles~\cite{wu2018physics} as well as by the amount and type of available data. The following are general guidelines based on prevailing practices:
 
 $\bullet$  {\em Local non-dimensionalization: }  To ensure applicability across different problems, the features should be locally non-dimensionalized~\cite{tracey2015machine,ling2016machine}. For instance, if the strain-rate tensor $\tilde{S}$ is used as a feature,  non-dimensionalization with respect to a local time scale (e.g.$\tilde{S}k/\epsilon$)  offers a greater possibility to generalize across different configurations, when compared to global non-dimensionalization.
 
 $\bullet$ {\em Invariance considerations: } As in traditional turbulence modeling~\cite{spalart2015philosophies}, ideal features should satisfy rotational, reflectional and frame-invariance properties. This aspect is discussed in detail in ~\cite{ling2016machine,wu2018physics}. This should apply to both the selected features and variables used for local non-dimensionalization.
 
 $\bullet$ {\em Local vs non-local features: } From the viewpoints of generalizability and implementation, it is desirable to have local features (such as  $\tilde{S}k/\epsilon$). In practice, wall-distance~\cite{wang2017physics-informed,wu2018physics} and wall-stress-based measures~\cite{singh2017machine-learning-augmented} appear to be important. Pressure-gradient-based features~\cite{wu2018physics} have also been used as a surrogate for non-local information.
 
 \begin{figure}
 	\centering
 	\includegraphics[width=0.8\textwidth]{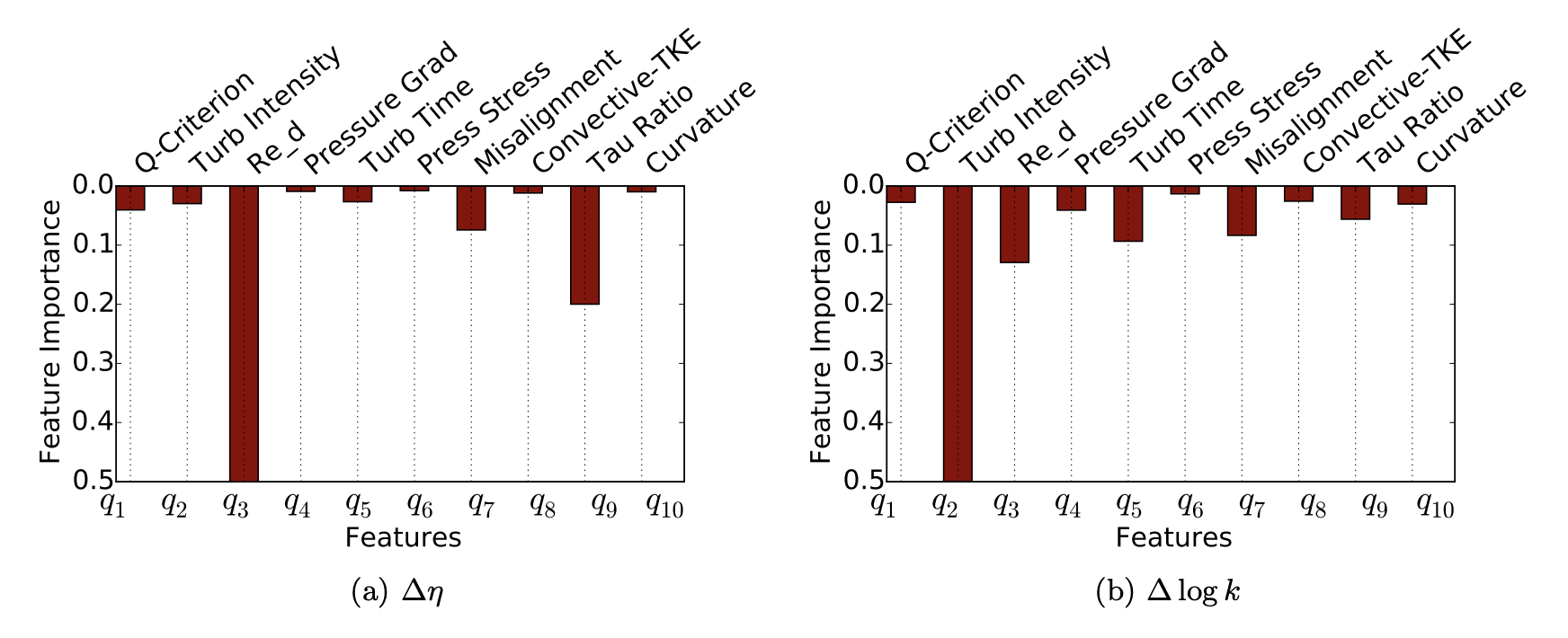}
 	\caption{Ranking important features using two different representations of model discrepancy(Ref.~\cite{wang2017physics-informed}) .}
 	\label{fig:features}
 \end{figure}
 
 $\bullet$ {\em Data considerations:}   Feature selection is a well-studied topic in the machine learning community and a comprehensive  summary can be found in Ref.~\cite{guyon2003introduction}.  Wang et al.~\cite{wang2017physics-informed} use random forest regression on 47 features and rank them in order of importance (Fig.~\ref{fig:features}). Such tools can be used as a guide to narrow down to a smaller set, with the caveat that limited data might bias the choice of features and impact model reliability. Indeed, constructing a model with $d$ features requires enough data to populate a $d$-dimensional feature space.

\section{Constraints}
\label{sec:constraints}
The general philosophy of combining ML models with physics-based models relies on the notion that physics information will complement the information contained in the data to yield generalizable predictions. The imposition of constraints can effectively reduce the search space of ML models to lower dimensional manifolds consistent with the physics.  In the present context of data-augmented turbulence modeling, constraints can take several forms, including:

$\bullet$ {\em Input constraints on the ML model:} As discussed in Section~\ref{sec:features}, the feature space $\tilde{\eta}_m$ can be constrained to a manifold that  satisfies invariance properties.

$\bullet$ {\em Output constraints on the ML model:}  The output of the learning model may also be constrained. For instance, if the output corresponds to a Reynolds stress perturbation to a baseline model (represented by a subscript b), i.e. $\tau_m = \tau_b + \delta(\tilde{\eta}_m;w)$ realizability constraints could be enforced on $\delta(\tilde{\eta}_m;w)$. Additional examples include the case in which ML outputs correspond to quantities that are reflectionally or rotationally invariant. This can be addressed - for instance - via data-augmentation~\cite{ling2016machine}.

$\bullet$ {\em Constraints on the outputs of the physics model:} In model-consistent training, relevant equality and inequality constraints can be imposed on observables of the output of the physics model. For instance, in a combustion modeling setting, even if the local heat release rate is not used in the loss function, the model output can be constrained to match the integrated heat release in the experiment (or DNS) during the training stage.

$\bullet$ {\em Constraint satisfaction via priors:} In the case of Bayesian inference, physical and mathematical information  can be used to enforce priors on the parameters and outputs.

Ref.~\cite{taghizadeh2020turbulence} also presents additional perspectives on compatibility and physics-based constraints.

\section{Additional Challenges \& Perspectives}
\label{sec:perspectives}
Turbulence modeling is a peculiar endeavor, with competing philosophies and paradoxes~\cite{duraisamy2017status}. 
For instance, it is often not clear as to what type of physical information - however elegant it may appear - is useful for predictive outcomes. In fact, Ref.~\cite{spalart2015philosophies} goes on to say that 
:  `` The central role of creativity and free intuition introduces a danger of {\em proliferation}. Any type of new term can be proposed, and many will satisfy the consensus constraints such as Galilean invariance, so that rejecting them becomes a matter of opposing intuition.'' It has to be recognized that ML augmentations are being introduced to a scenario which is already extremely complex, and sometimes counter-intuitive. Thus underlying intricacies and the Occam's Razor must be appreciated. 
We begin this section by outlining additional  challenges, and then offer some perspectives towards the development of generalizable ML-augmented models. 

{\em Irrecoverable model discrepancies}: Coarse-grained simulations of turbulence - as in the context of RANS and LES - have to contend with irrecoverable loss of information which is a consequence of the fact that, at any given instant in time, there are infinite realizations of velocity fields that are compatible with the coarse-grained field. Thus,  each of these realizations might evolve dynamically in different ways, which can only be captured in an average sense in a coarse-grained model~\cite{langford1999optimal,mishra2016sensitivity}. Further, the modeling structure by itself  (e.g. Single-point closure~\cite{duraisamy2019turbulence}, Markovian closures~\cite{parish2017non}, etc.) may introduce errors not addressable by the ML augmentation. 

\begin{figure}
	\centering
	\includegraphics[width=0.7\textwidth]{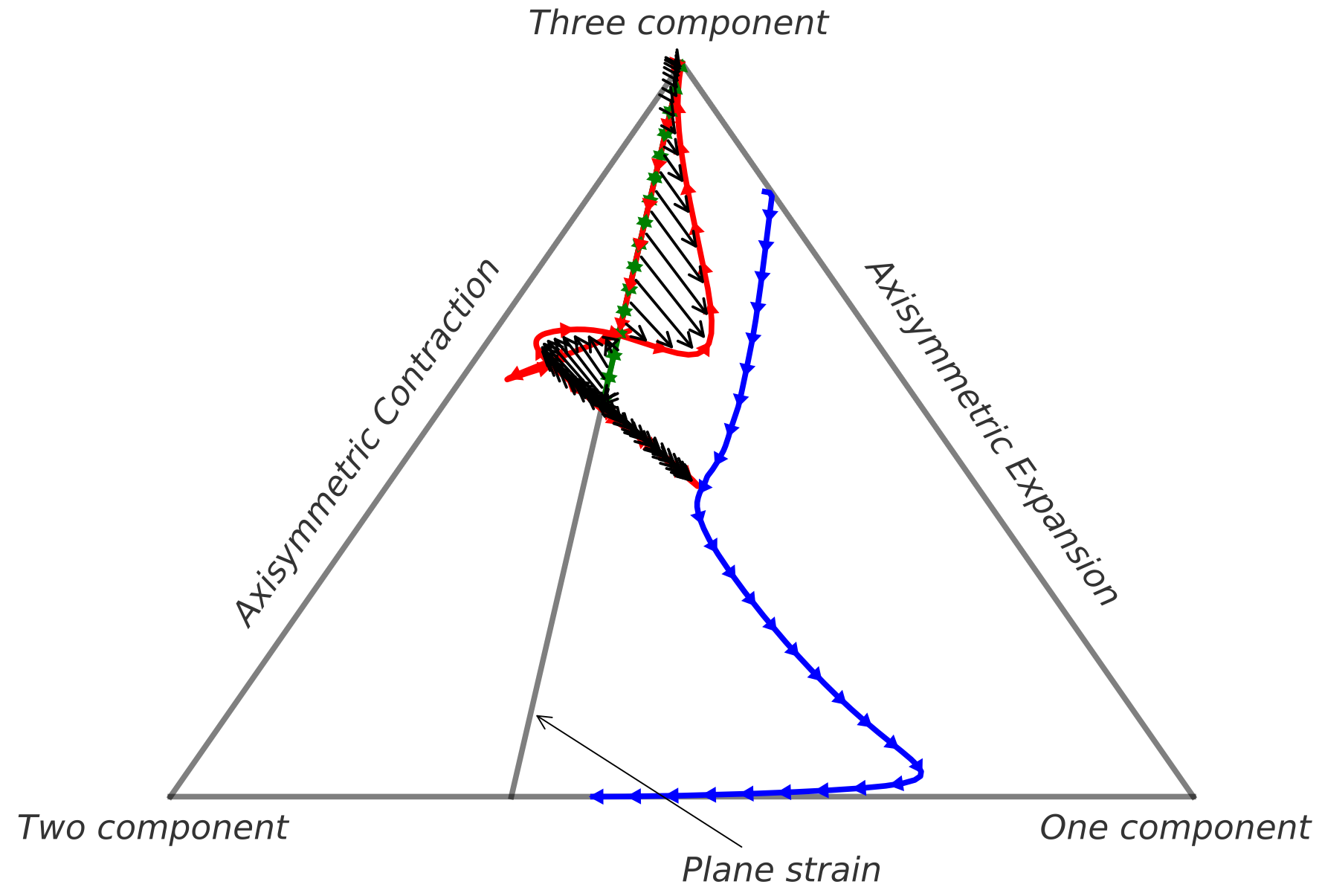}
	\caption{Model-consistent inference based on velocity data in Channel flow. Blue: Anisotropy predicted by DNS; Red: Anisotropy predicted by the inferred model. }
	\label{fig:anis}
\end{figure}

{\em Identifiability:} As a consequence of the fact that a) the model inadequacy cannot be fully addressed by the ML augmentation alone, and b) the available data may not be sufficient to learn the true model form, the augmentation may not be identifiable. Figure~\ref{fig:anis} illustrates the latter aspect. A model-consistent perturbation to the Reynolds stress anisotropy tensor based on the mean velocity was found to yield highly accurate mean velocity and Reynolds shear stress predictions, but the anisotropy was not predicted accurately. This could indeed be remedied if the entire Reynolds stress tensor was used as the data.   This example, however, is used a proxy for the lack of data for practical configurations:  DNS would be infeasible, and experimental methods will offer sparse measurements of certain field quantities.

{\em Impact of discretization error and filtering:} The augmentation of LES models is complicated by the fact that the resolved scales are seldom free of numerical errors, and thus disambiguating numerical errors from modeling errors is a challenge. This is especially true when explicit filtering~\cite{lund2003use,mathew2006new} is not employed. Thus, ascertaining what exactly has to be represented by the ML algorithm is typically unclear. Model-consistent training can account for the impact of numerical errors in a more precise manner, at least as far as  reconstructing the training data is concerned.  A further complication in LES is that the discrepancy between the filtered Navier-Stokes equations and the model equations may be difficult to assess. This is especially true when the filter is implicit, and model-consistent training as presented in Section~\ref{sec:consistency} requires further refinement. Further, as pointed out by Lund~\cite{lund2003use}, even explicit filtering may not fully address these deficiencies. 
Much further research is required towards the end of developing techniques for generalizable ML-augmented LES models.

{\em Interpretability:}  Interpretable ML models are desirable from the viewpoint of analysis, implementation, reproducibility, and wider use.  Models based on symbolic regression~\cite{weatheritt2016novel,haghiri2020large} are appealing in this regard.  Ref.~\cite{du2019techniques} offers a survey of recent research on the interpretability of a broader range of ML models.  It should be mentioned that while interpretable models are desirable, they do not automatically become applicable beyond the scenarios in which they were trained.

As with the development of all ML-based models, careful consideration should be given to achieving predictive capabilities beyond the training set.  While the topics discussed in this paper present some directions towards achieving generalization, the author's perspective is that ML-augmentation should be considered as just another tool in turbulence modeling. Judicious use of supervised and unsupervised learning, grounded by physics constraints  and mathematical rigor {\em and an understanding the information (and lack thereof) contained in the available data} is required  to derive generalizable models. 

\begin{figure}
	\centering
	\includegraphics[width=0.7\textwidth]{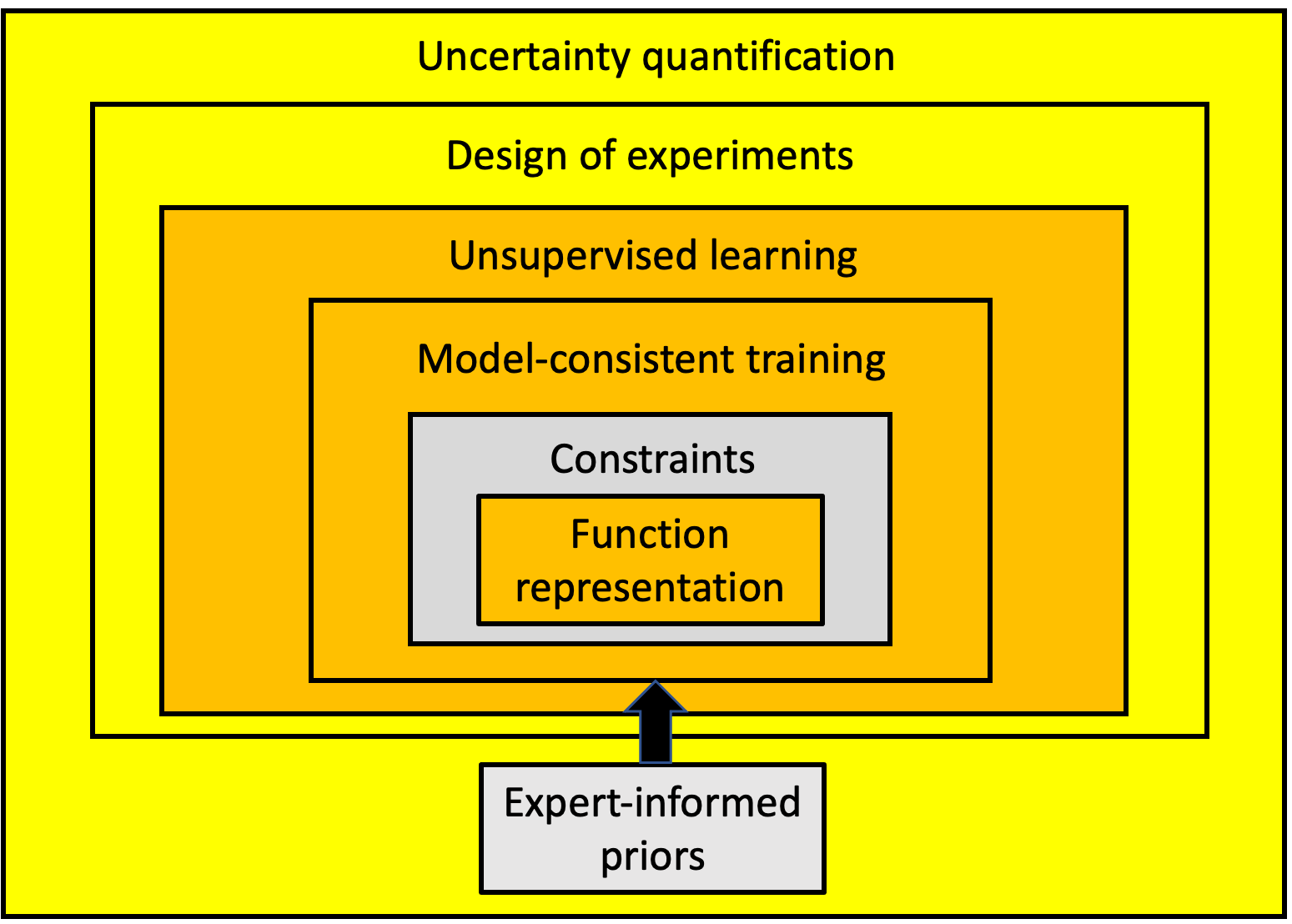}
	\caption{Schematic representing key layers in the  development of generalizable ML-augmented models.}
	\label{fig:comp}
\end{figure}

A schematic of a comprehensive approach towards developing ML-augmented models is shown in Figure~\ref{fig:comp}. Since every step of the modeling process involves assumptions, quantifying the impact of these assumptions in the form of distributions, and formally accounting for them in inverse and forward modeling will add robustness to model development. Over the past decade, uncertainty quantification formalisms have been applied to turbulence models~\cite{oliver2011bayesian,jofre2018les,xiao2019quantification}. While the treatment of parametric uncertainties is well-established and is purely a matter of applying scalable computational algorithms, addressing model form uncertainties continues to present an open  topic in computational science.
 
It is the author's opinion that the promise of ML-augmented turbulence modeling is clear. Preliminary successes have been demonstrated by small groups of individuals in a limited set of problem configurations. Several researchers have shown that ML-augmented modeling can offer improved predictions over classical models in problems that were either part of the training dataset or in problems that are related to the training set (e.g. an airfoil with a slightly different shape with respect to the training datasets~\cite{singh2017machine-learning-augmented}). The goal of more generalizable models has, however, not been achieved. Further progress requires the community to move beyond the publication of journal articles, and establish standards/benchmarks and foster  a collaborative ecosystem. ML (and artificial intelligence in general) have revolutionized certain fields rapidly, mainly because of a) wide use of real-world benchmarks (e.g. ImageNet), b) standardizing evaluation and accessibility of underlying tools, and c) promoting reproducibility of research.

As  a final point, throughout the history of turbulence (and combustion) modeling, the desire for deterministic models has resulted in developers presenting ``unique'' model parameters and model forms, after balancing many considerations. Documentation of candidate model forms that were explored during model development and specification of parameter ranges can help inform rigorous priors and greatly benefit data-driven modeling.

\section*{Acknowledgements}

The author acknowledges support from the NSF CBET program (\#1507928), ONR Sub-surface Hydrodynamics program  (\#N00014-17-1-2200), and NASA TTT program (\#80NSSC18M0149).   Mr. Aniruddhe Pradhan's help with the super-resolution results is gratefully acknowledged.




\bibliography{citations}{}
\bibliographystyle{aiaa}

\end{document}